\documentclass[14pt]{extarticle}
\usepackage[utf8]{inputenc}
\usepackage[T1]{fontenc}
\usepackage{bibentry}
\usepackage{amsmath}
\usepackage{amsfonts}
\usepackage{amssymb}
\usepackage{wrapfig}
\usepackage{graphicx}
\begin{document}
\begin{center}
 \today \\
\begin{Large} {\bf 
Relativistic deceleration vs acceleration, Unruh effect observation, and the Schott energy.}
\end{Large}
\vspace{10mm} \\
\begin{Large}
{Yefim S. Levin }\\
Department of  Mathematics, Salem State University,
Salem MA, 01970   \\
\end{Large}
\end{center} 
\begin{abstract}	
	The purpose of this article is to examine finite-time relativistic deceleration and the corresponding energy balance within the Lorentz-Abraham-Dirac equation, with special attention to boundary contributions associated with the Schott energy.
	
	From both experimental and kinematical points of view, deceleration differs significantly from acceleration. The large proper accelerations or decelerations required for possible observational effects, for example in connection with the Unruh effect, may be more naturally realized in deceleration scenarios than in comparable acceleration scenarios. Moreover, deceleration from a finite initial energy to rest can occur only over a finite time interval, in sharp contrast with idealized acceleration, which can continue indefinitely. 
	Thus, any finite episode of acceleration or deceleration must include the boundary transitions in the radiation energy balance.
	
	We first obtain explicit expressions for the time and distance required for a relativistic particle with nonzero initial velocity to come to rest under constant proper deceleration. When applied as a consistency check to the estimate of Lynch, Cohen, Hadad, and Kaminer (LCHK), these expressions show that the extremely large quoted deceleration should not be interpreted as a sustained classical uniform proper deceleration over a macroscopic crystal length. Under such an interpretation, the stopping time and stopping distance would be far too small. The corresponding upper bound on a sustained uniform deceleration is, according to our calculation, many orders of magnitude below the LCHK estimate.
	
	We then consider a straight-line trajectory of a charged particle entering and leaving an interval of uniform acceleration or deceleration, without velocity jump. We show that this motion can be described within the Lorentz-Abraham-Dirac (LAD) equation if the external force includes impulsive contributions during the short transition intervals between inertial and uniformly accelerated or decelerated motion. In this formulation, the external work is not locally converted directly into radiation during the interval of uniform acceleration or deceleration. Instead, the Schott energy acts as an intermediate reservoir: it is changed by the external work during the transition intervals and supplies the energy radiated during the subsequent uniform-acceleration or uniform-deceleration interval. 
	This property- idealized LAD radiation mechanism—is absent both from the Lorentz-equation description and from any LAD description that does not properly account for the boundary conditions. Finally, we offer a speculative interpretation of the Schott term as a classical forerunner of a quantum radiation process.
 \end{abstract} 
\section{Introduction}
At first sight, a comparison between uniform relativistic acceleration and uniform relativistic deceleration may seem unnecessary. A hyperbolic trajectory considered over the entire interval $(-\infty,\infty)$ contains both parts: for negative times it may be viewed as a decelerating motion with unphysical infinite initial energy, while for positive times it becomes an accelerating motion. Moreover, the Lorentz-Abraham-Dirac (LAD) equation has been discussed for many decades from many different perspectives; reviews of the related difficulties may be found, for example, in \cite{1960 Fulton Rohrlich,1969-Ginzburg,2010-Hammond, 2026_Yaghjian, 2011 Piazza}. It is therefore natural to ask whether anything useful can still be learned from this classical problem.
The renewed interest in the Unruh effect and in possible experimental tests of acceleration-induced thermality \cite{2021_Lynch},
\cite{2025 Levin} gives a reason to revisit the issue. The important
 point for the present paper is not only the magnitude of an acceleration parameter, but the time interval over which such an acceleration or deceleration can be sustained. At the Stanford Linac, for example, an accelerating field of $10\,\mathrm{MV/m}$ corresponds to a proper acceleration of about $2\times10^{17}\,\mathrm{m/s^2}$ and to an Unruh temperature of about $0.7\times10^{-3}\,\mathrm{K}$ \cite{2007-Crispino}. Electric fields of order $3.5\times10^9\,\mathrm{V/m}$, still far below the Schwinger limit $E\sim10^{18}\,\mathrm{V/m}$, correspond to an electron acceleration of order $7\times10^{19}\,\mathrm{m/s^2}$ and to an Unruh temperature of order $0.47\,\mathrm{K}$. In contrast, estimates connected with channeling experiments have quoted a positron deceleration as large as $5\times10^{39}\,\mathrm{m/s^2}$, corresponding formally to an enormous Unruh temperature \cite{2021_Lynch}.

A finite-energy particle, however, cannot decelerate uniformly forever. Deceleration to rest necessarily occurs over a finite time and a finite distance. This elementary observation provides a useful consistency check on any interpretation of very large deceleration estimates. In particular, if a very large instantaneous momentum loss is reinterpreted as a sustained classical uniform proper deceleration, the resulting stopping distance may become so small that the interpretation cannot be compatible with a millimeter-scale channeling experiment.

A second finite-time issue appears in the classical radiation problem
\footnote{This issue is also a subject of investigation related with Unruh effect  \cite{2006 Louko} }. When a uniformly accelerated or decelerated segment has a beginning and an end, the transitions between inertial and non-inertial motion affect the energy balance. We use the idealized model considered by Sokolov and Ternov \cite{1986-Sokolov} as a starting point. Their technique is useful, but we argue that the transition terms must be displayed explicitly in order to clarify the role of the Schott energy and to avoid an ambiguous interpretation of the external work and the radiated energy.
 
 \indent The paper is organized as follows. Section 2 reviews the kinematics of constant proper acceleration and deceleration, with emphasis on the finite stopping time and stopping distance for deceleration from a finite initial energy. Section 3 applies the LAD energy balance to a finite interval of uniform acceleration or deceleration, including the transition intervals, the role of the Schott term, and the connection with the Lorentz finite-size model of the electron. Section 4 compares this finite-time description with Gr\o n's no-preacceleration solution. Section \ref{Speculation} is more speculative: it asks whether the transition-force picture and the Schott-energy balance can be interpreted as a classical precursor of an internal-energy mechanism for classical radiation. This interpretation is separated from the main derivation. Section 6 summarizes the main conclusions, and the Appendix comments on related interpretations of energy balance in uniform acceleration/deceleration.

\section{Constant  proper acceleration vs deceleration}
\label{Constant proper acceleration}
For simplicity, we first consider the motion of a particle in two-dimensional Minkowski space. Its coordinates, velocity, and rate of change of velocity (that is, acceleration/deceleration) are
\begin{eqnarray}
	\label{coordinates velocity}
x^i\equiv (x^0,x^1)=(ct,x^1), \;\;\;
U^i\equiv (U^0,U^1)=(\frac{dx^0}{ds},\frac{dx^1}{ds})=\left( \frac{1}{\sqrt{1-\frac{v^2}{c^2}}}, \frac{v}{c\sqrt{1-\frac{v^2}{c^2}}}
\right),
\end{eqnarray}
\begin{eqnarray}
\frac{d U^i}{d s}\equiv \left( \frac{dU^0}{ds},\frac{dU^1}{ds} \right) =
\left( \dot{v}\frac{v}{c^2(1-\frac{v^2}{c^2})^{3/2}}, \;\;\;\dot{v} \frac{1}{c(1-\frac{v^2}{c^2})^{3/2}}\right), \;\;\;  
\dot{v}=\frac{dv}{ds}, \;\;\; s=c \tau
\end{eqnarray}
where $s$ is the invariant interval and $\tau$ is the proper time of the particle. The point of interest here is a motion when the scalar product
\begin{eqnarray}
	\frac{d U^i}{ds}\frac{dU_i}{ds}=-\frac{\dot{v}^2}{c^2}\frac{1}{(1-\frac{v^2}{c^2})^2}
\end{eqnarray}
does not depend on $\tau$ and is the same for any instantaneous inertial proper reference frame of the particle \cite{1973_Landau}
\begin{eqnarray}
	\frac{\dot{v}^2}{c^2}\frac{1}{(1-\frac{v^2}{c^2})^2}=w^2=const >0
\end{eqnarray}
or
\begin{eqnarray}
	\label{dif equation}
	\frac{\dot{\beta}}{1-\beta^2}	= w, \;\;\; \dot{\beta}\equiv\frac{1}{c}\frac{dv}{ds}.  
\end{eqnarray}
It is easy to see that
\begin{eqnarray}
w=\dot{\beta}_{v=0}	=\frac{1}{c}\frac{dv}{d s}\mid_{v=0} =\frac{1}{c^2}\frac{dv}{d \tau}\mid_{v=0}=\pm \frac{1}{c^2}a, \;\;\; a>0.
\end{eqnarray}
where $a$ is a constant proper acceleration (or deceleration).
For a particle moving in the positive direction of the $x$-axis and for positive $\tau$, with $\beta>0$, the plus sign in (\ref{dif equation}) corresponds to acceleration, since $\beta$ and $\dot{\beta}$ have the same sign. Conversely, the minus sign corresponds to deceleration, since $\beta$ and $\dot{\beta}$ have opposite signs.\\
\indent  Solution of differential equation (\ref{dif equation}) is:
\begin{eqnarray}
	ws=\frac{1}{2} \ln\frac{1+\beta}{1-\beta} +\frac{1}{2}\ln\varkappa
\end{eqnarray}
and
\begin{eqnarray} 
	\label{beta eqn}
	\beta(\tau)=\frac{e^{2wc \tau}-\varkappa}{e^{2wc\tau}+\varkappa},
\end{eqnarray}
where $\varkappa$ is a constant of integration. It significantly depends on an initial condition. We consider two of them. \\
\indent The initial condition with $\beta_{0}\equiv\beta|_{\tau=0}=0$ results in velocity of a hyperbolic motion along $x$-axis with a constant proper acceleration a (not deceleration!) for $\tau>0$:
\begin{eqnarray}
v(\tau)=c \tanh(cw\tau)=c \tanh(\frac{a}{c}\tau), \;\;\; \varkappa=1
\end{eqnarray} 
and can be found in any textbook.\\
\indent The case with a nonzero initial condition 
\begin{eqnarray}
	\frac{v_0}{c}\equiv\beta_0 >0
	\end{eqnarray}
is less commonly emphasized in the present context.
In this case, from (\ref{beta eqn}) we have
\begin{eqnarray}
	\beta_0=\frac{1-\varkappa}{1+\varkappa}, \;\;\;  \varkappa=\frac{1-\beta_0}{1+\beta_0}, \;\;\;
	\beta(\tau)=\frac{e^{2wc\tau}-\frac{1-\beta_0}{1+\beta_0}}{e^{2wc\tau}+\frac{1-\beta_0}{1+\beta_0}},
\end{eqnarray}
or  setting 
\begin{eqnarray}
\label{the sign of variable $w$}
w=-\frac{a}{c^2}, \;\;\; a  > 0
\end{eqnarray}
and introducing a new variable $\xi$
\begin{eqnarray}
	\label{full deceleration proper time}
	\frac{1-\beta_0}{1+\beta_0}=e^{-2a\xi /c}, \;\;\; \xi=\frac{c}{2a}\ln\frac{1+\beta_0}{1-\beta_0}>0,
\end{eqnarray}
we obtain
\begin{eqnarray}
	\label{beta on tau}
\beta(\tau)=\tanh(\frac{a}{c} (\xi-\tau)) > 0
\end{eqnarray}
It is easy to see that the velocity along the positive $x$-axis is decreasing and that $\xi$ is the proper time required for a complete stop, when $\beta(\xi)=0$.  \\
\indent Below we present some useful formulas for relativistic motion with a uniform deceleration.
From (\ref{beta on tau}) and (\ref{coordinates velocity}) we get the Lorentz factor in an instantaneous inertial frame co-moving along the $x$-axis
\begin{eqnarray}
	\label{gamma}
	\gamma(\tau)=\frac{1}{\sqrt{1-\beta^2(\tau)}}= \cosh\frac{a}{c}(\xi-\tau),
\end{eqnarray}
 four-velocity of a decelerating particle in an inertial lab (fixed) system 
\begin{eqnarray}
	\label{four velocity}
	U^{\mu}(\tau)=\frac{1}{c}\frac{dx^{\mu}}{d \tau}=\left(\cosh(\frac{a}{c}(\xi-\tau)), \;\sinh(\frac{a}{c}(\xi-\tau)),\;0,\;0\right),\;\;\; \mu=0,1,2,3, 
\end{eqnarray}
and its four-deceleration  
\begin{eqnarray}
	\label{four deceleration}
	\dot{U^{\mu}}=\frac{d U^{\mu}}{d s}=\frac{1}{c}\frac{d U^{\mu}}{d \tau} =
	\left(-\frac{a}{c^2}\sinh\frac{a}{c}(\xi -\tau), \;-\frac{a}{c^2}\cosh\frac{a}{c}(\xi -\tau),\;0, \;0  \right)
\end{eqnarray}
Particle coordinates, after integration of $u^{\mu}(\tau)$ over $\tau$ with initial conditions
\begin{eqnarray}
	t_{\tau=0}=0, \;\;\; x^k_{\tau=0} =0, \;\;\; k=1,2,3, \;\;\; 0 \le\tau \le \xi 
\end{eqnarray}
are
\begin{eqnarray}
	\label{lab coordinates}
	x^{\mu}(\tau)=(ct, x^1,x^2,x^3)= \nonumber \\
	\left(\frac{c^2}{a}[\sinh\frac{a}{c}\xi-\sinh\frac{a}{c}(\xi -\tau)], \frac{c^2}{a}[\cosh\frac{a}{c}\xi-\cosh\frac{a}{c}(\xi -\tau)],0,0 \right), 
\end{eqnarray}
where
\begin{eqnarray}
	0 < t \le t_{stop}, \;\;\; 0\le x^1(\tau) \le x_{stop}, \nonumber \\
t_{stop}=t(\tau=\xi)= \frac{c}{a}\sinh\frac{a}{c}\xi , \;\;\;  x_{stop}=x^1(\tau=\xi)= \frac{c^2}{a}\left(\cosh\frac{a}{c}\xi-1 \right),
\end{eqnarray}
Here $t_{stop}$ and $x_{stop}$ are, respectively, the lab-frame time of motion and the lab-frame distance traveled by the particle under constant proper deceleration with a finite initial velocity before it comes to a complete stop. \\
\indent To compare this with an experiment, it is convenient to represent these quantities with the help of 
(	\ref{beta on tau}, \ref{gamma}) in terms of initial velocity, $\beta_0$,  and initial Lorentz factor, $\gamma_0$ 
\footnote{In nonrelativistic case, $\beta_0 \ll 1$ and $at \ll c$, they have correct classical expressions
	\begin{eqnarray}
		t_{stop}=\frac{v_0}{a},\;\;\;x_{stop}=\frac{v_0^2}{2 a}.
\end{eqnarray} }:
\begin{eqnarray}
\label{stop time distance}
t_{stop}=\frac{c}{a}\frac{\beta_0}{\sqrt{1-\beta_0^2}}=\frac{c}{a}\sqrt{\gamma_0^2-1}, \nonumber \\
x_{stop}=\frac{c^2}{a}\left( \frac{1}{\sqrt{1-\beta_0^2}}-1\right)=\frac{c^2}{a}(\gamma_0-1)
\end{eqnarray}
\indent 
Excluding proper time $\tau$ from (\ref{lab coordinates}) we  get
\begin{eqnarray}
\left(\frac{c^2}{a}\right)^2= 
\left(x^1-\frac{c^2}{a} \cosh\frac{a}{c}\xi\right)^2 -\left( ct-\frac{c^2}{a}\sinh\frac{a}{c}\xi
\right)^2, \nonumber \\
0\le t \le t_{stop}=\frac{c}{a}\sinh\frac{a\xi}{c}
\end{eqnarray}
Then the coordinate $x^1(t)$, written in terms of the lab time and satisfying the initial conditions $x^1(0)=0$ and 
$\gamma(t=\tau=0)=\gamma_0$, is
\footnote{In nonrelativistic case, $at \ll c$ and $\beta_0 \ll 1$, it has correct expression
$x^1(t)\approx c\beta_0t -\frac{1}{2}at^2$
}
\begin{eqnarray}
x^1(t)=\frac{c^2}{a}\left(\cosh\frac{a}{c}\xi -\sqrt{1+(\frac{at}{c}-\sinh \frac{a\xi}{c})^2}
\right )= \nonumber \\
\frac{c^2}{a}\left( \gamma_0-\sqrt{1+\left(\frac{a t}{c}-\sqrt{\gamma_0^2-1}\right)^2}
\right)
\end{eqnarray} 
and  similarly
\begin{eqnarray}
\beta(t)=\frac{\sinh\frac{a}{c}(\xi-\tau)}{\cosh \frac{a}{c}(\xi-\tau)}=
\frac{-\frac{a}{c}t+\sinh\frac{a}{c}\xi}{-\frac{a}{c^2}x^1+\cosh\frac{a}{c}\xi}=
\frac{\sqrt{\gamma_0^2-1}-\frac{a}{c}t}
{\sqrt{1+(\sqrt{\gamma_0^2-1}-\frac{a}{c}t)^2}}\\
\gamma(t)=\sqrt{1+(\sqrt{\gamma_0^2-1} -\frac{a}{c}t)^2}
\end{eqnarray}
\subsection{Application to channeling deceleration estimates}
The distinction between acceleration and deceleration becomes especially relevant in discussions of radiation reaction and acceleration-induced thermality in aligned-crystal channeling experiments. In \cite{2018_Wistisen}, ultrarelativistic positrons were modeled semiclassically: the particles propagate through an aligned crystal according to a classical Lorentz-force trajectory, while photon emission is treated probabilistically with quantum energy and probability distributions. In this description the photon emission is accompanied by a positron momentum loss.

A related idea was used in \cite{2021_Lynch}, where the positron momentum loss was interpreted in terms of deceleration. According to their estimate, a positron with energy $178.2\,\mathrm{GeV}$ and Lorentz factor $\gamma=(3.5)10^5$ may be associated with a proper deceleration of order $a=(5)10^{39}\,\mathrm{m/s^2}$ during emission of a photon with characteristic energy $\hbar\omega_0=150\,\mathrm{GeV}$.The corresponding emission time scale is quoted in \cite{2021_Lynch} (above (S79)) as \footnote{In the present paper this number is used only as the quoted short momentum-transfer time scale; the notation $\omega_0$ should be read consistently with the convention used in \cite{2021_Lynch}.}
\begin{eqnarray}
\Delta t= \frac{\pi}{\omega_0}=(8.6)10^{-26}\,\mathrm{s}.
\end{eqnarray}
Thus the quoted value is naturally associated with an extremely short momentum-change interval, not with a sustained uniform deceleration through the crystal. This distinction is important because uniform acceleration is the condition normally used in the standard Unruh-effect argument \cite{1982_Birrell}.

For the purpose of a consistency check, let us nevertheless ask what would follow if the same numerical value were interpreted as a classical uniform proper deceleration. Using formulas (\ref{stop time distance}), a positron with the same initial Lorentz factor and deceleration, $a=(5)10^{39}\,\mathrm{m/s^2}$, would come to rest after
\begin{eqnarray}
t_{stop}=2.1 \times 10^{-26}\,\mathrm{s}
\end{eqnarray}
having traveled only
\begin{eqnarray}
	x_{stop}=6.3 \times 10^{-16}\,\mathrm{cm}.
\end{eqnarray}
This shows that LCHK deceleration parameter, $a=(5)10^{39}\,\mathrm{m/s^2}$, should not be reinterpreted as sustained classical proper deceleration through the $3.8$ or $10\,\mathrm{mm}$ crystal widths used in the experimental setup of \cite{2018_Wistisen}.

There is also a more basic limitation on the classical interpretation. The above $t_{stop}$ and $x_{stop}$ are far below characteristic atomic scales: $x_{stop}$ is much smaller than the Bohr radius, and $t_{stop}\ll T_{Bohr}=1.5\times10^{-16}\,\mathrm{s}$, where $T_{Bohr}$ is the electron revolution period in the Bohr model of the hydrogen atom. Therefore the estimate $a=(5)10^{39}\,\mathrm{m/s^2}$ should not be understood as a sustained classical uniform proper deceleration. At most, it may characterize an extremely short momentum-transfer event in a semiclassical or quantum-radiative description.
\subsubsection{Consistency bound for sustained uniform deceleration}
Let us use (\ref{stop time distance}) to estimate the largest sustained uniform deceleration compatible with a particle traversing a macroscopic channeling length. If a classical uniform deceleration is to describe motion through the crystal, then the stopping distance should exceed the relevant dechanneling length $L_d$. For the experimental setup of \cite{2018_Wistisen}, where crystals of widths $d=3.8$ and $10\,\mathrm{mm}$ were used, taking $L_d$ to be at least of order $10\,\mathrm{mm}$ gives the conservative estimate
\begin{eqnarray}
a=\frac{c^2(\gamma_0-1)}{x_{stop}} <\frac{c^2(\gamma_0-1)}{d}=3.15 \times 10^{24}\,\mathrm{m/s^2}.
\end{eqnarray}
This value is many orders of magnitude smaller than the deceleration estimate quoted in \cite{2021_Lynch}. The corresponding Unruh temperature,  about $(1.3)10^4\,\mathrm{K}$, is also far below the formally quoted value $(2)10^{19}\,\mathrm{K}$. The conclusion is not that the channeling experiment cannot involve large instantaneous momentum changes, but rather that such changes should not be reinterpreted as a sustained classical uniform proper deceleration over the crystal length.
Deceleration $3.15 \times 10^{24}\,\mathrm{m/s^2}$ should therefore not be regarded as a definite experimental bound on sustained uniform deceleration. Rather, it should be understood as an order-of-magnitude consistency bound for the experimental setup.
\section{ Uniform acceleration/deceleration in a finite time interval }
\label{Uniform acceleration in a finite time interval}
\subsection{External work for finite-time uniform acceleration/deceleration}
The preceding discussion emphasizes that decelerated motion has a definite beginning and end. This makes the boundary conditions at the transitions between inertial motion and uniformly accelerated or decelerated motion central to the radiation problem. We therefore consider a charge moving along a prescribed straight-line trajectory with four-velocity 
\begin{eqnarray}
\label{trajectory}	
u^{\mu}= \frac{d x^{\mu}}{d \tau}\equiv \dot{x}^{\mu}=
\begin{cases}
(c \cosh\eta_i,	0,0, c \sinh\eta_i ),& \text{if $\tau< 0$;}\\
(c \cosh\eta(\tau),	0,0, c \sinh\eta(\tau) ),&\text{$0<\tau<\tau_f$;} \\
(c \cosh\eta_f,	0,0, c \sinh\eta_f ), & \text{$\tau >\tau_f$}
\end{cases},	 
\end{eqnarray}
where 
\begin{eqnarray}
\label{rapidity}	
\eta(\tau)=\eta_i +s\frac{a}{c}\tau, s=\pm 1, \;\;\; \eta_i=\eta(0^+), \;\;\;
 \eta_f = \eta(\tau_f^-), \;\;\; 0<\tau <\tau_f.
\end{eqnarray}
Here $\tau$ is the proper time, $c$ is the speed of light, and $a$ is the magnitude of the constant proper acceleration or deceleration.
$s=+1$ corresponds to acceleration and $s=-1$ to deceleration \footnote{Do not confuse it with an invariant interval s used in Section\ref{Constant proper acceleration}.}. The Lorentz factor is
$\gamma(\tau)=\cosh \eta(\tau)$ and $\gamma_f=\cosh\eta_f, \gamma_i=\cosh\eta_i$. \\
\indent The corresponding 4-acceleration is a discontinuous function \footnote{This motion has neither pre-acceleration nor runaway tendency. } 
\begin{eqnarray}
a^{\mu}(\tau)=\frac{d u^{\mu}}{d \tau}\equiv\dot{u}^{\mu} =(\dot{u}^0,\dot{u}^1,\dot{u}^2,\dot{u}^3)=
(\theta(\tau)-\theta(\tau-\tau_f)) \times a^{\mu}_{hyp}(\tau), \nonumber \\
a^{\mu}_{hyp}(\tau)=s a(\sinh\eta(\tau),0,0, \cosh \eta(\tau)) \nonumber \\
\end{eqnarray}
\begin{eqnarray}
\theta(\tau)=
\begin{cases}
	1 ,& \text{if $\tau >0$;}\\
	0,& \text{if $\tau <0$.}
\end{cases}	 
\end{eqnarray}
We use the metric signature $(+,-,-,-)$, so that $a_\nu a^\nu=-a^2<0$ for constant proper acceleration. Therefore $\dot{a}^{\mu}=\frac{d a^{\mu}}{d \tau}$ contains delta-function contributions:
\begin{eqnarray}
\label{acceleration change rate}	
\dot{a}^{\mu}(\tau)=(\theta(\tau)-\theta(\tau-\tau_f)) \dot{a}^{\mu}_{hyp}(\tau)
+a^{\mu}_{hyp}(0^+)\delta(\tau)-a^{\mu}_{hyp}(\tau_f^-)\delta(\tau-\tau_f), \nonumber \\
\dot{a}^{\mu}_{hyp}=\frac{a^2}{c^2}u^{\mu}(\tau)=-\frac{a_{\nu}a^{\nu}}{c^2}u^{\mu}(\tau), \;\;\; a=const,
\end{eqnarray}
which correspond to acceleration jump transitions between constant-velocity motion and uniform acceleration/deceleration at $\tau=0$ and $\tau=\tau_f$. The delta functions should be understood as the idealized limit of a switching process whose duration is very short compared with $\tau_f$ \footnote{More details can be found in Section "Generalization of some properties of continuous functions to the delta-function "\cite{1986-Sokolov}}, not as a literal claim that a physical force can be discontinuous. \\
\indent  We assume that this prescribed motion is generated by an external force and ask what force is required by the Lorentz-Abraham-Dirac (LAD) equation \footnote{A related approach to a similar problem is discussed in Sec. \ref{Gron comparison}.}
\begin{eqnarray}
\label{LAD}
f^{\mu}_{ext}=ma^{\mu}- \Gamma^{\mu}.
\end{eqnarray}
Abraham 4-vector, $\Gamma^{\mu}$, can be represented as  \cite{Rohrich}(6-55)
\begin{eqnarray}
\Gamma^{\mu}=\frac{2e^2}{3c^3}(\dot{a}^{\mu}+ \frac{a^{\nu}a_{\nu}}{c^2}u^{\mu})=\frac{2e^2}{3c^3}(\dot{a}^{\mu}- \frac{a^2}{c^2}u^{\mu}),
\;\;\; g^{\mu \nu}=(+,-,-,-)
\end{eqnarray}
or \cite{2004 Eriksen Gron} (Section 7)
\begin{eqnarray}
\label{radiation reaction force}	
\Gamma^{\mu}=-\dot{p}^{\mu}_s-\dot{p}^{\mu}_{rad}, 
\end{eqnarray}
where 
\begin{eqnarray}
\label{schott 4 momentum}
\dot{p}^{\mu}_s=- \frac{2 e^2}{3 c^3}\dot{a}^{\mu}, \nonumber \\
\dot{p}^{\mu}_{rad}= \frac{2 e^2}{3 c^3}(\frac{a^2}{c^2} u^{\mu}),
\end{eqnarray}
and 
 $p^{\mu}_s$ and $p^{\mu}_{rad}$ denote, respectively, the Schott 4-momentum and radiation field 4-momentum. Then LAD equation (\ref{LAD}) in terms of energy-momentum changes is
 \begin{eqnarray}
 \label{external force}	
 f^{\mu}_{ext}=m a^{\mu}+\dot{p}^{\mu}_s+\dot{p}^{\mu}_{rad}.
 \end{eqnarray}
\indent Because of the Schott term $\dot{p}_s^{\mu}(\tau)$, the supporting force \footnote{This force is also analyzed in Sec. \ref{Gron comparison}.}
contains singular transition contributions 
\begin{eqnarray}
(f^{\mu}_{ext})_{sing}=(\dot{p}^{\mu}_s)_{sing}  \nonumber
\end{eqnarray}
at $\tau=0$ and $\tau=\tau_f$. These terms have a direct effect on the system's energy balance
 \footnote{ Of course, as Hammond says, \cite{2010-Hammond}  "$\ddot{v}^{\mu}$ term also signals another problem.If the force is discontinuous, then $\ddot{v}^{\mu}$ is singular. Real forces are not discontinuous " but he says further "many examples use this artifice." The simple idealized model used here nevertheless makes important features of the LAD equation visible. About necessity of impulsive external force to get rid of pre-acceleration was also pointed out in \cite{2026_Yaghjian, 2006_Yaghjian}  }.\\
\indent The work done by this external force and changes in the energies of the particle and its electromagnetic field on different integration segments of the trajectory
are defined by the time component, $\mu=0$, of the equation
\begin{eqnarray}
\label{time component}
c f^0_{ext}= \dot{E}^{kin}+\dot{E}_s +\dot{E}_{rad},
\end{eqnarray} 
where 
\begin{eqnarray}
\dot{E}^{kin} \equiv \frac{d E^{kin}}{d \tau}=m c a^0, \\
\dot{E}_s =c  \dot{p}^0_s,    \\
\dot{E}_{rad}=c  \dot{p}^0_{rad}
\end{eqnarray}
denote the proper-time rates of change of kinetic, Schott, and radiation energies, respectively.\\
\indent There are two types of integration intervals. The first type consists of small neighborhoods of the transition points, (-$\epsilon, \epsilon$) and ($\tau_f-\epsilon,\tau_f+\epsilon$) with $\epsilon>0$, chosen such that 
\begin{eqnarray}
\int_{-\epsilon}^{\epsilon} \delta(\tau)d\tau=1, \;\;\;\int_{\tau_f-\epsilon}^{\tau_f+\epsilon} \delta(\tau-\tau_f)d\tau=1.\nonumber
\end{eqnarray}
These intervals are associated with singular parts of the Schott term $f_s^{\mu}$
and are especially important for our consideration because they describe boundary processes between inertial and uniform acceleration/deceleration modes  of motion\footnote{Integrals of $\delta$-functions over finite intervals are considered in \cite{1986-Sokolov}}. The second type consists of the intervals $(-\infty,0)$, $(0,\tau_f)$, and $(\tau_f,\infty)$, which include only non-singular contributions.\\
\indent It is easy to see that on the time segments $(-\infty,0)$ and $(\tau_f,\infty)$ no energy changes occur and then the external force work is also zero:
\begin{eqnarray}
W(-\infty,0)\equiv \int_{-\infty}^{0}c f^0_{ext}d\tau=\Delta E^{kin}(-\infty,0) +\Delta E_s(-\infty,0)+\Delta E_{rad}(-\infty,0)=0,  \nonumber \\
W(\tau_f,\infty)=0 \nonumber \\
\end{eqnarray}
because 
\begin{eqnarray}
	\Delta E^{kin}(-\infty,0)=\Delta E_s(-\infty,0)=\Delta E_{rad}(-\infty,0)=0
\end{eqnarray}
and similarly for segment $(\tau_f,\infty)$. \\
\indent On interval $(0,\tau_f)$ all external force work goes into the kinetic energy change 
\begin{eqnarray}
W(0,\tau_f) =\Delta E^{kin}(0,\tau_f),
\end{eqnarray}
where \begin{eqnarray}
\Delta E^{kin}(0,\tau_f)=mc \int_{0}^{\tau_f}a^0_{hyp}d\tau =
mc^2(\cosh \eta_f-\cosh \eta_i)=
mc^2(\gamma_f-\gamma_i), \nonumber \\
\end{eqnarray}
because in this time interval for a uniform acceleration/deceleration the radiated energy is balanced by a decrease of Schott energy
\begin{eqnarray}
	\dot{E}_s + \dot{E}_{rad}=0
\end{eqnarray} 
and corresponding integral is
\begin{eqnarray}
	\Delta E_s(0,\tau_f) +\Delta E_{rad}(0,\tau_f)=0,
\end{eqnarray}
It is worth noting that in both cases, deceleration with $s=-1, \eta_f< \eta_i$ and acceleration with $s=+1, \eta_f>\eta_i$,
$\Delta E_{rad}(0,\tau_f)>0$ and $\Delta E_s(0,\tau_f)<0$:
\begin{eqnarray}
	\Delta E_s(0,\tau_f) =-\Delta E_{rad}(0,\tau_f)=
	-\frac{2e^2}{3c^2}(s a)(\sinh \eta_f-\sinh \eta_i)
\end{eqnarray}
\indent This bookkeeping makes explicit a feature of the LAD energy balance related to the transitions at
$\tau=0$ and $\tau=\tau_f$
 between inertial and acceleration/deceleration modes. They are associated with 
 arbitrary small neighborhoods  $(-\epsilon,+\epsilon)$ and  $(\tau_f-\epsilon \tau_f+\epsilon)$ respectively:
\begin{eqnarray}
W(-\epsilon,\epsilon)=\Delta E_s(-\epsilon,\epsilon)=-\frac{2e^2}{3 c^2}a^0_{hyp}(0^+)=
-\frac{2e^2}{3c^2}(s a) \sinh \eta_i, \nonumber \\
W(\tau_f-\epsilon,\tau_f+\epsilon)=\Delta E_s(\tau_f-\epsilon, \tau_f+\epsilon)=\frac{2e^2}{3 c^2}a^0_{hyp}(\tau_f^-)=\frac{2e^2}{3c^2}(s a) \sinh \eta_f,
\end{eqnarray}
where the following definitions are used
\begin{eqnarray}
\Delta E_s(-\epsilon,\epsilon)=
-\frac{2 e^2}{3c^2} \int_{-\epsilon}^{\epsilon} a^0_{hyp}(0^+)\delta(\tau)d\tau \nonumber \\
\Delta E_s(\tau_f-\epsilon,\tau_f+\epsilon)=
+\frac{2 e^2}{3c^2} \int_{\tau_f-\epsilon}^{\tau_f+\epsilon} a^0_{hyp}(\tau_f^-)\delta(\tau-\tau_f)d\tau 
\end{eqnarray}
During these impulsive acceleration transitions, while the rapidity remains continuous, the work done by the external force changes only the Schott energy\footnote{Here we closely follow \cite{1986-Sokolov}, but this feature is not made explicit there. }. Kinetic energy, which is described by non-singular functions,
 does not change during these transitions and no radiation loss takes place:
\begin{eqnarray}
\Delta E^{kin}(0^-, 0^+)=\Delta E^{kin}(\tau_f^-,\tau_f^+ )=0,
\nonumber \\
 \Delta E_{rad}(0^-,0^+)=\Delta E_{rad}(\tau_f^-,\tau_f^+)=0.
\end{eqnarray}
\indent Thus, in our idealized model, the main finite-time LAD energy-balance result is that the total external work decomposes into two contributions: the kinetic-energy change during the uniform acceleration/deceleration interval and the Schott-energy changes during the two transition intervals:
\begin{eqnarray}
\label{radiation mechanism}
W=\Delta E^{kin} (0,\tau_f)+\Delta E_s(-\epsilon,+\epsilon)+\Delta E_s(\tau_f-\epsilon, \tau_f+\epsilon).
\end{eqnarray}
In this decomposition the external work is not converted directly into radiation during the uniform acceleration/deceleration interval. The radiated energy is supplied by the Schott-energy decrease during that interval, while the external force changes the Schott energy at the boundaries. \\
\indent To illustrate the role of the Schott energy in the radiation process, let us compare how the Schott energy changes during the transition intervals for two cases, acceleration and deceleration,
with the same magnitude of proper acceleration/deceleration $a$, with $\tau \in (0,\tau_f)$, and such that
\begin{eqnarray}
	\gamma^{acc}_i=\gamma^{dec}_f, \;\;\; \gamma^{acc}_f=\gamma^{dec}_i.
\end{eqnarray}
It means that 
\begin{eqnarray}
	\cosh\eta^{acc}_i= \cosh\eta_f^{dec}, \;\;\;
	\cosh\eta_f^{acc}=\cosh \eta^{dec}_i
\end{eqnarray}
\indent We will see that the Schott energy changes for acceleration and for deceleration during transitions at $\tau=0$ and $\tau=\tau_f$ are different. 
General definition of the Schott energy and momentum in 4-form \cite{2020_Kirk}(33):
\begin{eqnarray}
	p^{\mu}_s=-\frac{2e^2}{3c^3}a^{\mu}=(E_s/c,\vec{p}_s)
\end{eqnarray}
and the energy is
\begin{eqnarray}
	E_s=-\frac{2e^2}{3c^2}a^0
\end{eqnarray}
Then for acceleration and deceleration respectively we obtain
\begin{eqnarray}
\label{signs 1}	
	\Delta E_s^{acc}(-\epsilon,+\epsilon)=-\frac{2 e^2}{3 c^2}a\sinh\eta_i^{acc} \le 0
	, \nonumber \\
	\Delta E_s^{acc}(\tau_f-\epsilon,\tau_f+\epsilon)=+\frac{2 e^2}{3 c^2}a\sinh(\eta_i^{acc}+\frac{a}{c} \tau_f) > 0 
\end{eqnarray}
and 
\begin{eqnarray}
\label{signs 2}	
	\Delta E_s^{dec}(-\epsilon,+\epsilon)=\frac{2 e^2}{3c^2}a \sinh\eta_i^{dec} >0,
	\nonumber \\
	\Delta E_s^{dec}(\tau_f-\epsilon,\tau_f+\epsilon)=-\frac{2 e^2}{3c^2}a \sinh(\eta_i^{dec}-\frac{a}{c}\tau_f) \le 0.
\end{eqnarray}
Although the signs of $\Delta E_s$ at each transition are different for acceleration and deceleration, as can be seen by comparing the first and second equations in (\ref{signs 1}) with those in (\ref{signs 2}), the sums of the Schott-energy changes at the two transitions are equal:
\begin{eqnarray}
	\Delta E_s^{acc}(-\epsilon,+\epsilon)+\Delta E_s^{acc}(\tau_f-\epsilon,\tau_f+\epsilon) \nonumber \\
	=\Delta E_s^{dec}(-\epsilon,+\epsilon)+
	\Delta E_s^{dec}(\tau_f-\epsilon,\tau_f+\epsilon).
\end{eqnarray} 
\indent These results clarify a point that is often obscured when the transition intervals are not displayed explicitly. Their omission may introduce ambiguity into the interpretation of the energy balance. For example, in \cite{1986-Sokolov}, for a similar problem with $\eta(\tau)=\frac{a}{c}\tau$, $\eta_i=0$, and $s=+1$, the transition intervals between uniform acceleration and constant-velocity motion are not displayed explicitly. After (11.93) we read:
"...while the external force is acting $(0<\tau<\tau_0)$, the energy loss 
\footnote{They define energy loss rate $\dot{E}^{loss}$ as a sum of $\dot{E}_{rad}$ and $\dot{E}_s$ which they call respectively irreversible and reversible losses.}is zero (emission is compensated by the advanced effect), whereas $E^{rad}\ne 0$ during this time." \\
\indent A few lines below, after (11.95), they say:
"the overall energy balance obtains after application of all the forces $(\tau>\tau_0)$: the work done by the external force is equal to the increase 
\footnote{It corresponds in our notation to s=1 }
in the kinetic energy of the electron and the energy emitted by the electron during the time interval $\tau_0$ during which it is accelerated ".  From the first statement, it follows that radiation during the uniform  acceleration interval is balanced by the "advanced effect" \footnote{in other words, the Schott energy loss} and no external work was used
but from the second statement we find out that the radiation is supplied by a part of the work done by the external force during acceleration interval $(0,\tau_0)$\footnote{$\tau_0$ in our notations is $\tau_f$}. \\
\indent These two statements can be reconciled, but only if the Schott-energy changes at the beginning and end of the acceleration interval are kept explicitly. \\
\indent There is a reason for this confusion in \cite{1986-Sokolov} because the sum of
the Schott energy changes during impulsive transitions at $\tau=0$ and $\tau=\tau_f$ is equal to energy radiated during acceleration/deceleration interval $(0,\tau_f)$:
\begin{eqnarray}
\Delta E_s (-\epsilon,\epsilon) + \Delta E_s(\tau_f-\epsilon,\tau_f+\epsilon )=-\Delta E_s(0,\tau_f)=
\Delta E_{rad}(0,\tau_f)
\end{eqnarray}
and then
\begin{eqnarray}
	\label{energy conservation}
W=\Delta E^{kin}(0,\tau_f)+\Delta E_{rad}(0,\tau_f).
\end{eqnarray}
 Although (\ref{radiation mechanism}) and (\ref{energy conservation}) are formally equivalent, they emphasize different physics. Equation (\ref{radiation mechanism}) displays the intermediate role of the Schott energy. First, the impulsive external force at $\tau=0$ changes the Schott energy by $\Delta E_s(-\epsilon,\epsilon)$. Second, during the interval $(0,\tau_f)$ the Schott energy decreases and supplies the radiated energy, so that $\Delta E_s(0,\tau_f)=-\Delta E_{rad}(0,\tau_f)$. Third, the impulsive external force at $\tau=\tau_f$ changes the Schott energy by $\Delta E_s(\tau_f-\epsilon,\tau_f+\epsilon)$, returning it to the value appropriate to inertial motion. In this sense, for this idealized uniformly accelerated or decelerated segment, the Schott energy acts as a mediator between the external force and radiation.

\indent Computing only the total Schott-energy change on an interval $(\tau_1,\tau_2)$ with $\tau_1<0$, $\tau_2>\tau_f$, and $a(\tau_1)=a(\tau_2)=0$ hides the separate boundary contributions at $\tau=0$ and $\tau=\tau_f$. Related interpretations in the literature are discussed in the Appendix. \\

\subsection{Summary of the idealized LAD radiation mechanism}

The following table separates the nonsingular contribution from the uniform-acceleration/deceleration interval from the singular contributions associated with the two transition intervals.
\begin{table}[h]
\begin{center}
\scriptsize
\resizebox{\textwidth}{!}{%
\begin{tabular}{|c|l|c|c|c|c|}
\hline
Row $i$ & Interval $I$ & External work $W(I)$ & Schott-energy change $\Delta E_s(I)$& Radiated energy $\Delta E_{rad}(I)$& Kinetic-energy change $\Delta E^{kin}(I)$\\ \hline
1 & $(-\epsilon,\epsilon)$, singular transition & $-\frac{2 e^2}{3 c^2}sa\sinh \eta_i$ & $-\frac{2 e^2}{3 c^2}sa\sinh \eta_i$ & $0$ & $0$ \\ \hline
2 & $(0,\tau_f)$, nonsingular uniform interval & $mc^2(\gamma_f-\gamma_i)$ & $-\frac{2 e^2}{3 c^2}sa(\sinh\eta_f-\sinh \eta_i)$ & $\frac{2 e^2}{3 c^2}sa(\sinh\eta_f-\sinh \eta_i)$ & $mc^2(\gamma_f-\gamma_i)$ \\ \hline
3 & $(\tau_f-\epsilon,\tau_f+\epsilon)$, singular transition & $\frac{2e^2}{3c^2}sa\sinh\eta_f$ & $\frac{2e^2}{3c^2}sa\sinh\eta_f$ & $0$ & $0$ \\ \hline
&&1&2&3&4 \\ \hline
\end{tabular}
}
\caption{Energy changes on the singular transition intervals and on the nonsingular uniform-acceleration/deceleration interval  for a charge moving according to Eq.~(\ref{trajectory}).}
\label{tab:singular-nonsingular-energy}
\end{center}
\end{table}

For later reference, denote each entry in the four physical-energy columns by its table coordinate $(i,k), i=1,2,3; k=1,2,3,4.$  The table gives three complementary balance statements.

First, on the nonsingular interval $(0,\tau_f)$,
\begin{align}
\label{energy conservation 1}
(2,1)&=(2,4)
\quad\Rightarrow\quad
W(0,\tau_f)=\Delta E^{kin}(0,\tau_f),\\
\label{energy conservation 2}
-(2,2)&=(2,3)
\quad\Rightarrow\quad
-\Delta E_s(0,\tau_f)=\Delta E_{rad}(0,\tau_f).
\end{align}
These two relations describe the energy conservation  during the uniform part of the motion: the external work changes the kinetic energy, while the radiated energy is balanced by the decrease of Schott energy.

Second, the singular contributions at the two transition intervals satisfy
\begin{align}
\label{energy conservation 3}
(1,1)+(3,1)&=(1,2)+(3,2)\nonumber\\
\Rightarrow\quad
W(-\epsilon,\epsilon)&+W(\tau_f-\epsilon,\tau_f+\epsilon)\nonumber\\
&=\Delta E_s(-\epsilon,\epsilon)+\Delta E_s(\tau_f-\epsilon,\tau_f+\epsilon).
\end{align}
Thus, the singular part of the external work changes the Schott energy at the two boundaries.

Third, the Schott-energy changes over the two singular transition intervals and over the nonsingular uniform interval sum to zero:
\begin{align}
\label{energy conservation 4}
(1,2)+(3,2)+(2,2)&=0\nonumber\\
\Rightarrow\quad
\Delta E_s(-\epsilon,\epsilon)&+\Delta E_s(\tau_f-\epsilon,\tau_f+\epsilon)\nonumber\\
&+\Delta E_s(0,\tau_f)=0.
\end{align}
This identity makes explicit that, in this idealized bookkeeping, the Schott-energy decrease during the uniform interval is compensated by the boundary contributions.

For compact notation, define
\begin{align}
W_{sing}&\equiv W(-\epsilon,\epsilon)+W(\tau_f-\epsilon,\tau_f+\epsilon),\\
(\Delta E_s)_{sing}&\equiv \Delta E_s(-\epsilon,\epsilon)+\Delta E_s(\tau_f-\epsilon,\tau_f+\epsilon),\\
(\Delta E_s)_{nonsing}&\equiv \Delta E_s(0,\tau_f),\\
W_{nonsing}&\equiv W(0,\tau_f), \\
W_{total}& \equiv W_{nonsing}+W_{sing}
\end{align}
Equation (\ref{energy conservation 1}) alone also holds for motion described by the Lorentz equation when self-reaction is neglected. In that reduced description, however, the source of the radiated energy is not accounted for. The pair of relations (\ref{energy conservation 1}) and (\ref{energy conservation 2}) summarizes the energy conservation during the uniform part of the motion, but it does not by itself explain how the Schott energy is prepared at the beginning and restored at the end. The specific feature of the present idealized LAD model is contained in Eqs.~(\ref{energy conservation 3}) and (\ref{energy conservation 4}): the singular transition terms change the Schott energy at the boundaries. 

The central singular/nonsingular relation, a distinctive feature of this model, can then be written as
\begin{eqnarray}
\label{for speculation}
W_{sing}=(\Delta E_s)_{sing}=-(\Delta E_s)_{nonsing}=
\Delta E_{rad}(0,\tau_f).
\end{eqnarray}
This chain of equalities should be read from left to right: the singular part of the external work changes the Schott energy at the transition intervals, while the corresponding decrease of Schott energy during the uniform interval supplies the radiated energy.Thus the radiation process can be described as a sequence of energy conversions mediated by the Schott term and triggered by the impulsive transition forces.

This property—the idealized LAD radiation mechanism—is absent both from the Lorentz-equation description and from any LAD description that does not properly account for the boundary conditions.

The corresponding total work over an interval containing both transition regions and the uniform segment is usually written in the form \cite{1986-Sokolov}
\begin{align}
	W_{total}& =\Delta E^{kin}(0,\tau_f)+\Delta E_{rad}(0,\tau_f).
\end{align}

This compact expression, however, hides the boundary-Schott mechanism responsible for the energy transfer in the idealized LAD description.
\subsection{The LAD equation and the Lorentz model of electron}
Transitions between inertial motion and accelerated motion were also investigated by Yaghjian.  In his treatment, based on a reexamined Lorentz model of the electron, a charge enters a parallel-plate capacitor without pre-acceleration and with zero initial rapidity.  Yaghjian writes the result in terms of the rapidity $\nu(\tau)$ \cite{2026_Yaghjian}(A.3):
 \begin{eqnarray}
 \frac{\nu^{\prime}(\tau)}{c}=\frac{eE_0}{mc}(h(\tau)-h(\tau-\tau_2))+\frac{\Delta \nu_1}{c}\delta(\tau)+\frac{\Delta \nu_2}{c}\delta(\tau-\tau_2), \\
 \frac{\nu(\tau)}{c}=\frac{e E_0}{mc} \{\tau (h(\tau)-h(\tau-\tau_2) ) +\tau_2 h(\tau-\tau_2)\} +\frac{\Delta \nu_1}{c}h(\tau)+\frac{\Delta \nu_2}{c}h(\tau-\tau_2).
 \end{eqnarray}
\indent The corresponding equations in the notation used in the present paper would be
 \begin{eqnarray}
 \dot{\eta}(\tau)=\frac{sa}{c}(\theta(\tau)-\theta(\tau-\tau_f)) + \nonumber \\
(\eta(0^+)-\eta(0^-)) \delta(\tau) + (\eta(\tau_f^+)-\eta(\tau_f^-)) \delta(\tau-\tau_f)
 \end{eqnarray}
 and 
 \begin{eqnarray}
\eta(\tau)=\eta_i+\frac{sa}{c} \{ \tau (\theta(\tau)-\theta(\tau-\tau_f)) + \tau_f \theta(\tau-\tau_f) \}+\nonumber \\
(\eta(0^+)-\eta(0^-)) \theta(\tau) +(\eta(\tau_f^+)-\eta(\tau_f^-))\theta(\tau-\tau_f).
 \end{eqnarray}
with $\eta_i=0$ and $s=+1$, because Yaghjian considers acceleration rather than deceleration and uses zero initial rapidity \cite{2026_Yaghjian}(just after (5b)).  The important difference is the boundary prescription.  In Yaghjian's approach the rapidity jumps $\Delta\nu_1$ and $\Delta\nu_2$ may be non-zero.  He also points out that such jumps can lead to physically problematic features, including negative radiated energy across the second transition interval \cite{2026_Yaghjian}(A.3c,A.5c) and a possible incorrect split of the radiation/recoil force  into temporal and spatial components that do not transform as a Lorentz four-vector \cite{2026_Yaghjian}(discussion after (37)). \\
\indent The model used in this paper makes a different choice.  The rapidity history (\ref{rapidity}) is continuous at both transition points:
 \begin{eqnarray}
 \eta(0^-)=\eta(0^+), \qquad \eta(\tau_f^-)=\eta(\tau_f^+).
 \end{eqnarray}
Therefore the delta-function terms in $\dot\eta(\tau)$ vanish.  Nevertheless, $\ddot{\eta}(\tau)$ contains the delta-function term and the supporting external force is not a simple rectangular force.  Because the acceleration changes discontinuously at $\tau=0$ and $\tau=\tau_f$, $\dot a^{\mu}$ contains delta-function terms; through the Schott part of the LAD equation, Eq.~(\ref{schott 4 momentum}, \ref{external force}), these terms require impulsive contributions to the external force.  These force impulses change the Schott energy during the transition intervals, but they do not produce jumps of rapidity or velocity. \\
\indent The comparison with Yaghjian may therefore be stated explicitly as follows.  Yaghjian's construction allows rapidity/velocities jumps at the boundary transitions.  The present model allows the external force to contain the idealized impulses required by the LAD equation but keeps rapidity continuous.  Consequently, the rapidity-jump problems discussed by Yaghjian do not arise here; the price is that the boundary impulses in the external force must be kept explicitly in the energy-balance analysis.

\section{Comparison with Gr\o n's no-preacceleration solution}
\label{Gron comparison}
This section compares the finite-time uniformly accelerated/decelerated motion used in the present paper with the no-preacceleration solution discussed by Gr\o n.  The essential point is that both calculations use the same scalar LAD equation.  The difference is the way he uses it.  Gr\o n uses a rectangular force that is switched on and off without the boundary terms and obtains
the rapidity history. In the present paper the rapidity history is prescribed first, and the external force required by the LAD equation is then obtained from that history.  This complete force contains the rectangular part during the interval of uniform acceleration/deceleration and two transition impulses at the endpoints.  Substitution of this complete force into Gr\o n's integral formula is therefore a self-consistency check: it reproduces the prescribed finite-time rapidity and introduces neither preacceleration nor runaway behavior.

\indent For rectilinear motion it is convenient to use rapidity.  Gr\o n denotes rapidity by $\alpha$; here it is denoted by $\eta$.  The characteristic LAD time is
\begin{eqnarray}
\label{tau e definition Gron comparison}
\tau_e=\frac{2e^2}{3mc^3},
\end{eqnarray}
which is Gr\o n's $\tau_0$.  If $F(\tau)$ is the external force in the direction of motion, LAD  equation (24) \cite{2010 Gron} can be written in the notation of this paper as
\begin{eqnarray}
\label{Gron equation 24 in our notation}
\dot{\eta}-\tau_e\ddot{\eta}=\frac{F(\tau)}{mc}\equiv \Phi(\tau).
\end{eqnarray}
Here a dot denotes differentiation with respect to the proper time $\tau$.  Equation (\ref{Gron equation 24 in our notation}) is the common starting point for both comparisons below.

\indent The motion used in the present paper is (\ref{rapidity}) 
\begin{eqnarray}
\label{eta dot Gron comparison}
\dot{\eta}(\tau)=s\frac{a}{c}\left[\theta(\tau)-\theta(\tau-\tau_f)\right], \qquad s=\pm1,
\end{eqnarray}
and its distributional derivative is
\begin{eqnarray}
\label{eta double dot Gron comparison}
\ddot{\eta}(\tau)=s\frac{a}{c}\left[\delta(\tau)-\delta(\tau-\tau_f)\right].
\end{eqnarray}

\indent Substituting (\ref{eta dot Gron comparison}) and (\ref{eta double dot Gron comparison}) into (\ref{Gron equation 24 in our notation}) gives the normalized force required by the LAD equation for the prescribed finite-time motion:
\begin{eqnarray}
\label{our force in Gron equation}
\Phi_{our}(\tau)=s\frac{a}{c}\left[\theta(\tau)-\theta(\tau-\tau_f)\right]
-\tau_e s\frac{a}{c}\delta(\tau)
+\tau_e s\frac{a}{c}\delta(\tau-\tau_f).
\end{eqnarray}
Equivalently,
\begin{eqnarray}
\label{our physical force in Gron equation}
F_{our}(\tau)=msa\left[\theta(\tau)-\theta(\tau-\tau_f)\right]
-m\tau_e sa\delta(\tau)
+m\tau_e sa\delta(\tau-\tau_f).
\end{eqnarray}
The first term is the ordinary force acting during $0<\tau<\tau_f$.  The two delta-function terms are the idealized switch-on and switch-off impulses associated with the Schott term.  They are not an additional equation of motion; they are what Eq.~(\ref{Gron equation 24 in our notation}) requires when the rapidity derivative is made discontinuous at the endpoints.

\indent The same force was used in Section~\ref{Uniform acceleration in a finite time interval}, where it appears in four-vector form as $f^{\mu}_{ext}$.  In the present scalar form it makes the comparison with Gr\o n especially transparent.  The rectangular part of (\ref{our force in Gron equation}) agrees with the constant-force part of Gr\o n's example (\ref{Gron rectangular force comparison}) when $A=sa/c$. The only difference is the presence, in (\ref{our force in Gron equation}), of the two Schott transition impulses.

\indent The solution of (\ref{Gron equation 24 in our notation}) with the lower integration limit in the past is Gr\o n's Eq.~(28), written in our notation as
\begin{eqnarray}
\label{Gron equation 28 in our notation}
\dot{\eta}(\tau)=-\frac{e^{\tau/\tau_e}}{\tau_e}
\int_{-\infty}^{\tau}\Phi(\tau^{\prime})e^{-\tau^{\prime}/\tau_e}d\tau^{\prime}.
\end{eqnarray}
We now put the complete force (\ref{our force in Gron equation}) into the integral solution (\ref{Gron equation 28 in our notation}).  For $\tau<0$ the integral is zero, and hence
\begin{eqnarray}
\label{Gron comparison solution before}
\dot{\eta}(\tau)=0,\qquad \tau<0.
\end{eqnarray}
For $0<\tau<\tau_f$, one obtains
\begin{eqnarray}
\int_{-\infty}^{\tau}\Phi_{our}(\tau^{\prime})e^{-\tau^{\prime}/\tau_e}d\tau^{\prime}
=s\frac{a}{c}\tau_e\left(1-e^{-\tau/\tau_e}\right)-\tau_e s\frac{a}{c}
=-\tau_e s\frac{a}{c}e^{-\tau/\tau_e}.
\end{eqnarray}
Therefore
\begin{eqnarray}
\label{Gron comparison solution inside}
\dot{\eta}(\tau)=s\frac{a}{c},\qquad 0<\tau<\tau_f.
\end{eqnarray}
Thus the complete force reproduces the prescribed constant value of $\dot{\eta}$ during the finite interval.

\indent For $\tau>\tau_f$, the integral contains the rectangular contribution and both endpoint impulses:
\begin{eqnarray}
\int_{-\infty}^{\tau}\Phi_{our}(\tau^{\prime})e^{-\tau^{\prime}/\tau_e}d\tau^{\prime}
=s\frac{a}{c}\tau_e\left(1-e^{-\tau_f/\tau_e}\right)
-\tau_e s\frac{a}{c}
+\tau_e s\frac{a}{c}e^{-\tau_f/\tau_e}=0.
\end{eqnarray}
Therefore
\begin{eqnarray}
\label{Gron comparison solution after}
\dot{\eta}(\tau)=0,\qquad \tau>\tau_f.
\end{eqnarray}
Equations (\ref{Gron comparison solution before}), (\ref{Gron comparison solution inside}), and (\ref{Gron comparison solution after}) give exactly the derivative of the rapidity assumed in the finite-time model.  Integrating over $\tau$ gives (\ref{rapidity})	
Thus Gr\o n's integral formula, when supplied with the complete LAD force (\ref{our force in Gron equation}), reproduces the finite-time rapidity history of the present paper and contains no runaway term.

\indent For comparison, take the rectangular force used in Gr\o n's no-preacceleration example and write it in the present notation as
\begin{eqnarray}
\label{Gron rectangular force comparison}
\Phi_G(\tau)=A\left[\theta(\tau)-\theta(\tau-\tau_f)\right], \qquad A=\frac{s a}{c}.
\end{eqnarray}
This force has the same constant value as the rectangular part of (\ref{our force in Gron equation}), but it omits both Schott transition impulses.  Substitution into (\ref{Gron equation 28 in our notation}) gives, before the force is switched on,
\begin{eqnarray}
\dot{\eta}_G(\tau)=0,\qquad \eta_G(\tau)=\eta_i,\qquad \tau<0.
\end{eqnarray}
Thus this solution is also a no-preacceleration solution.

\indent During the interval $0<\tau<\tau_f$, the rectangular force gives
\begin{eqnarray}
\dot{\eta}_G(\tau)=A\left(1-e^{\tau/\tau_e}\right),
\end{eqnarray}
and hence
\begin{eqnarray}
\label{Gron interval rapidity comparison}
\eta_G(\tau)=\eta_i+A\tau-A\tau_e\left(e^{\tau/\tau_e}-1\right),
\qquad 0<\tau<\tau_f.
\end{eqnarray}
After the force is switched off, one obtains
\begin{eqnarray}
\dot{\eta}_G(\tau)=-A\left(e^{\tau/\tau_e}-e^{(\tau-\tau_f)/\tau_e}\right),
\end{eqnarray}
and therefore
\begin{eqnarray}
\label{Gron after rapidity comparison}
\eta_G(\tau)=\eta_i+A\tau_f-A\tau_e\left(e^{\tau/\tau_e}-e^{(\tau-\tau_f)/\tau_e}\right),
\qquad \tau>\tau_f.
\end{eqnarray}
These are the same no-preacceleration expressions as Gr\o n's Eqs.~(29)--(31)\cite{2010 Gron}, written in the rapidity notation used here and shifted to the interval $0<\tau<\tau_f$.  The exponential term in (\ref{Gron after rapidity comparison}) persists after the external force has vanished; this is the runaway part of the no-preacceleration solution.

\indent The comparison may now be stated compactly.  With the rectangular force (\ref{Gron rectangular force comparison}), Gr\o n's integral formula (\ref{Gron equation 28 in our notation}) gives the no-preacceleration but runaway solution (\ref{Gron interval rapidity comparison})--(\ref{Gron after rapidity comparison}).  With the complete force (\ref{our force in Gron equation}), including the two Schott transition impulses, the same integral formula gives the finite-time rapidity (\ref{rapidity}).  Therefore the distinction is not a disagreement about the LAD equation.  It is a difference between two different external-force prescriptions.

\section{Speculative remarks about a mechanism of classical radiation}
\label{Speculation}
In the idealized model considered above, a finite interval of uniform acceleration or deceleration can be represented in the LAD equation only if the external force contains singular transition terms at the beginning and at the end of the interval. The calculation fixes the role of these terms in the energy balance, but it does not by itself explain their microscopic origin. \\
\indent A useful physical picture is provided by Yaghjian's charged-sphere derivation of the LAD equation. He writes \cite{2026_Yaghjian}, p.~88, that ``the necessity of a transition force $f_a(t)$, which contributes only during the short time it takes light to travel across the charged sphere, can be understood physically by considering the interaction of two differential elements of charge at either end of the charge distribution.'' For a sphere of finite radius, this transition force is associated with communication between different parts of the extended charge distribution and with the corresponding internal energy.  In the limiting procedure, as radius tends to zero, that leads to the LAD equation and it is tempting to keep the same interpretation of the transition force and the associated Schott energy as a manifestation inner an structure of a charge. But in this case, for the point-particle equation, one cannot literally speak about  interaction between different parts and a resolved internal structure of the electron. A more cautious statement is that the LAD transition force and corresponding Schott energy change are the remnants of respectively the finite-size transition force acting on the particle and its inner energy change, all located in a point-like particle. \\

\indent Teitelboim's splitting \cite{1970_1971_Teitelboim} gives a complementary way to say this. Away from the world line the radiative and bound parts of the electromagnetic energy-momentum tensor are separately conserved,
\begin{eqnarray}
	\partial_\mu T^{\mu\nu}_{rad}=0, \qquad \partial_\mu T^{\mu\nu}_{bound}=0,
\end{eqnarray}
so any exchange between the bound part, the particle, and the radiative part must be localized on the singular world line or, for a finite-size model, inside the small world tube surrounding the charge. In this distributional sense, the conversion of bound energy into radiation is attached to the charge, not to an extended region of empty space. \\
\indent Thus, within this idealized model, the LAD equation provides additional insight into the classical radiation as a local process described by set of equations (\ref{for speculation}).
 
\indent The picture suggests, and here our speculation starts, a limited analogy with quantum emission: Schott energy change term as a classical forerunner of a quantum radiation process because the outgoing radiation is associated with a change of a localized inner energy. The analogy should not be overstated. The LAD equation remains a classical equation, the Schott energy is not quantized here, and the present calculation does not by itself prove that classical radiation has a quantum character. It only suggests that the classical Schott term plays a role formally reminiscent of an internal energy participating in radiation.
\footnote{The idea of a quantum interpretation of the Schott energy was put forward in \cite{2019_Khokonov}.}

\section{Concluding remarks}
The main point of this paper is twofold. First, a very large deceleration inferred from a short momentum-transfer event should not automatically be interpreted as a sustained classical uniform proper deceleration. For a finite-energy particle, constant proper deceleration implies definite stopping times and distances; applying these formulas to the quoted channeling estimate \cite{2021_Lynch} leads to stopping scales of order $10^{-26}\,\mathrm{s}$ and $10^{-16}\,\mathrm{cm}$, which are not compatible with a macroscopic crystal-length interpretation.

Second, a finite interval of uniform acceleration or deceleration necessarily has transition regions. In the idealized LAD model considered here these regions are represented by impulsive force terms. Keeping these terms explicit clarifies the energy balance and energy conversions in time: during the uniform acceleration/deceleration interval the radiated energy is supplied by the decrease of the Schott energy, while the external force changes the Schott energy at the boundaries. So radiation is a process which can be described as a sequence of energy conversions mediated by the Schott term and triggered by the impulsive transition forces. This property—the idealized LAD radiation mechanism—is absent both from the Lorentz-equation description and from any LAD description that does not properly account for the boundary conditions.

Finally, the speculative interpretation developed above suggests that 
LAD radiation process is a classical forerunner of a quantum radiation process because it involves an inner, Schott, energy change.This last interpretation should be kept distinct from the main calculation: it is a classical analogy motivated by the LAD energy balance and by the finite-size charged-sphere picture, not a proof that the Schott energy is quantized.

\appendix
\section{Comments on energy balance in uniform acceleration/deceleration}
\indent Hammond's article \cite{2010-Hammond}, p.~227, contains the following excerpt from Sorkin's work \cite{2004_Sorkin}:
\begin{quote}
``A well-known peculiarity of the radiation reaction force on a charged particle is that it vanishes when the particle accelerates uniformly. But this raises a paradox. An accelerating charge radiates, and the longer the acceleration continues, the greater the total energy radiated. If one asks where this energy comes from in the case of uniform acceleration, the usual answer is that it is `borrowed' from the near field of the particle and then `paid back' when the acceleration finally ceases. But this `debt' can be arbitrarily great if the acceleration remains uniform for a long enough time. What, then, if the agent causing the acceleration decides not to repay the borrowed energy? What if, in fact, it does not even possess enough energy to pay its immense debt at that time? If we believe in conservation of energy, the respective answers must be that the accelerating agent must not be at liberty to avoid transferring the required energy and that it must always possess the necessary amount to cover its accumulated debt.''
\end{quote}
In the notation of this paper, the radiation-reaction force mentioned in this passage is $f^{\mu}_{rr}(\tau)$. The terminology of borrowing and repayment is striking, but the finite-time analysis above gives a more precise interpretation. Uniform acceleration or deceleration cannot be physically separated from the transition intervals at its beginning and end. In the present LAD model, these transitions are represented by the impulsive parts of the external force. They supply or compensate the changes in Schott energy and thereby remove the ambiguity about the source of the radiated energy.

\indent A second example is the discussion in \cite{2010_Gal'tsov}. On p.~2, Gal'tsov summarizes several features of uniformly accelerated motion. The following comments compare those statements with the finite-time LAD model developed in this paper.

\begin{description}
\item[D.G.] ``Therefore the energy-momentum balance of the system consisting of the accelerated charge and its Maxwell field includes three, but not just two, ingredients: the particle momentum, the momentum carried by radiation, and the bound electromagnetic momentum.''

\item[Y.L.] This statement is correct, provided that the work done by the external force is also included. In the finite-time formulation used here, the external force is especially important at the two transition intervals, where it changes the Schott, or bound-field, energy.

\item[D.G.] ``The radiation momentum can be extracted both from the particle momentum and indirectly from the bound momentum.''

\item[Y.L.] This conclusion is not the one obtained in the specific model considered above. During the interval of uniform acceleration or deceleration, the radiated energy is supplied by the decrease of the bound, or Schott, energy. It is not taken directly from the particle's kinetic energy, and it is not supplied directly by the external work during the uniform part of the motion.

\item[D.G.] ``This explains the origin of radiation of the uniformly accelerated charge, in which case the total reaction force is zero and thus the kinetic particle momentum is constant.''

\item[Y.L.] In a fixed inertial frame the kinetic momentum of the particle is not constant. It increases during acceleration and decreases during deceleration. What vanishes for uniform proper acceleration is the total LAD self-force, not the time derivative of the mechanical momentum in an inertial frame.

\item[D.G.] ``Physically, however, the acceleration has to start at some moment and to finish at some moment, and during the stages of acquiring and losing the acceleration the bound momentum is exchanged with the kinetic momentum.''

\item[Y.L.] In the finite-time model analyzed here, the transition intervals have a different role. At $\tau=0$ and $\tau=\tau_f$, the impulsive external force changes the Schott momentum. These boundary changes do not produce a velocity jump and therefore do not directly change the kinetic momentum. The kinetic momentum changes during the interval of uniform acceleration or deceleration, whereas the Schott-energy boundary changes are supplied or compensated by the external force at the endpoints.
\end{description}



\begin{thebibliography} {99}
 	\bibitem{1960 Fulton Rohrlich} T. Fulton, F. Rohrlich,  Classical Radiation from a Uniformly Accelerated Charge, Annals of Physics 9,499-517 (1960.) 
 	\bibitem{1969-Ginzburg} V.L. Ginzburg, Radiation and radiation friction force in uniformly accelerated motion, Usp. Fiz. Nauk, 97,569 (1969). 
 	\bibitem{2010-Hammond}	Richard T. Hammond, Relativistic Particle Motion and Radiation Reaction in Electrodynamics, Electronic Journal of Theoretical Physics ( EJTP) 7, No. 23 (2010) 221–258.  
 	\bibitem{2026_Yaghjian} A.D. Yaghjian, Conservation of momentum and energy in the Lorentz-Abraham-Dirac equation of motion, arXiv:2512.02960, 19 Jan. 2026. 
 	\bibitem{2011 Piazza} A. Di Piazza, C. Muller, K.Z. Hatsagorthyan,  C.H. Keitel, Extremely high-intensity laser interactions with fundamental quantum systems,  arXiv:1111.3886 v1 [hep-ph] 16 Nov 2011.  Review of Modern Physics, 84, 1177 (2012).
 	\bibitem{2021_Lynch} Lynch, M.H., Cohen, E., Hadad, Y.,Kaminer,I., Experimental Observation of Acceleration Induced Thermality, Phys. Rev. D 104, 025015 ( 2021 )  
 	\bibitem{2025 Levin} Y.S. Levin, Some theoretical aspects of observation of acceleration induced thermality, Phys. Rev D 111(6), 065021(2025)
 	\bibitem{2007-Crispino} Luis C.B. Crispino, A. Higuchi, G.E.A.
 	Matsas, The Unruh effect and its applications,  arXiv:0710.5373v1 [gr-qc] 29 Oct 2007.
 	Rev. Mod. Phys. 80,787 (2008).  
 	\bibitem{2006 Louko}Jorma Louko and Alejandro Satz, 
 	How often does the Unruh-DeWitt detector click? Regularisation by a spatial profile.
 	arXiv:gr-qc/0606067v3 10 Oct 2006
 	\bibitem{1986-Sokolov}A.A. Sokolov, I.M. Ternov, Radiation from Relativistic Electrons, New York 1986; A.A.Sokolov, I.M.Ternov, Relativistic Electron, Moscow 1974 (In Russian) 
 	\bibitem{1973_Landau} L.D.Landau, E.M.Lifshitz, The Field Theory (Russian), 1973
 	\bibitem{2018_Wistisen} T.N. Wistisen, A.D. Piazza, H.V. Knudsen, U.I. Uggerhoj. Experimental evidence of quantum radiation reaction in aligned crystals,  Nature Communications  (2018) 9:795 
 	\bibitem{1982_Birrell} N. D. Birrell, P. C.W. Davis, Quantum fields in curved space
 	 \bibitem{2004 Eriksen Gron} E. Eriksen and Ø. Grøn, Electrodynamics of hyperbolically accelerated charges V. The field of a charge in the Rindler space and the Milne space, Annals of Physics 313 (2004) 147–196 
 	 \bibitem{Rohrich} F.Rohrlich, Classical Charged Particles, Syracuse University, Addison-Wesley Publishing Company, Inc. Reading, Massachusetts
 	\bibitem{2006_Yaghjian} Arthur D. Yaghjian, Relativistic Dynamics of a Charged Sphere, 2nd ed.,Lect. Notes Phys. 686 (Springer, New York 2006), DOI 10.1007/b98846
 	\bibitem{2020_Kirk} Kirk T. McDonald, On the history of the radiation reaction, Joseph Henry Laboratories, Princeton University, Princeton, NJ 08544 (2020).
 	\bibitem{2004_Sorkin} R.D.Sorkin, An energy bound deduced from the vanishing of the radiation reaction force during uniform acceleration, Mod. Phys. Lett A19, 543 (2004).
 	\bibitem{2010_Gal'tsov}D. Gal'tsov, Radiation reaction and energy-momentum conservation,  arXiv:1012.2846v1 [gr-qc] 13 Dec 2010.
 	\bibitem{2010 Gron} Gron, The significance of the Schott energy for energy-momentum conservation of a radiating charge obeying the Lorentz-Abraham-Dirac equation, arXiv 2010; Am. J. Phys. 79, 115–122 (2011)
 	\bibitem{1970_1971_Teitelboim}C. Teitelboim, Splitting of the Maxwell Tensor: Radiation Reaction without Advanced Fields, Phys. Rev. D1,1572-1582 (1970). Phys. Rev. D2, 1763 (1970) 
 \bibitem{2019_Khokonov} M. Kh. Khokonov, On the quantum interpretation of the classical Schott term in the theory of radiation damping, Phys. Letters B 791, 281-286 (2019)
 \end{thebibliography}
\end{document}